\documentclass[journal=jacsat,manuscript=article]{achemso}

\usepackage{upgreek}
\usepackage{xcolor} 
\usepackage[version= 3]{mhchem} 
\usepackage{amsmath} 
\mciteErrorOnUnknownfalse

\author{Bidisha Bhatt}
\affiliation[University of Oslo]
{Mechanics Section, Department of Mathematics, University of Oslo, Norway}

\author{Andreas Carlson}
\affiliation[University of Oslo]
{Mechanics Section, Department of Mathematics, University of Oslo, Norway}
\alsoaffiliation[Ume{\aa} University] 
{Department of Medical Biochemistry and Biophysics, 901 87 Ume{\aa}, Sweden}
\email{acarlson@math.uio.no}

\title[An \textsf{achemso} demo]
  {Capillary Slinky: Equilibrium and Dynamics of Droplets in a Soft Spring}

\abbreviations{IR,NMR,UV}
\keywords{American Chemical Society, \LaTeX}

\begin{document}


\begin{abstract}
Springs can be found in many applications and biological systems, and when these are soft, they easily deform. At small scales, capillarity can induce a force leading to spring deformations when the elastocapillary number is small. We demonstrate through experiments the non-trivial equilibrium shape liquid droplets adopt in these soft springs, which form an annulus, Eruciform, and spherical shapes. When these droplets are set in motion, they display different flow regimes with significant dissipation generated by the internal rotational flow. The static and dynamics of droplets in such a capillary slinky is also used to demonstrate how surface tension can actuate springs by stretching/compression, while providing a way for active flow control in soft springs. 
\end{abstract}

\section{Introduction}
Helical-shaped fibers are found across a wide range of engineering applications, such as in textile designs,\cite{lu2024functional, xue2023soft, zhao2023hydrogel} humidity and flow sensors,\cite{li2012superelastic, aziz2023plant} fog harvesting\cite{shi2018fog, yang2025active}, packaging, \cite{yuan2023novel} water purification/emulsification, \cite{ jia2025application} and proposed as active materials to name but a few examples. One such soft realisation is a slinky, a classical toy made for kids to play with in stairs, which is illustrative of how these structures easily stretch, bend, and twist. Helical slender structures are also found across a wide range of biological systems as a fundamental structure \cite{chouaieb2006helices} illustrated by the DNA, \cite{calladine1997understanding} plant tendrils \cite{aziz2023plant}, bacterial flagella, \cite{berg2003rotary, zang2025dynamics} plant vascular \cite{wang2024capillary, sasaki2023confined, schneider2021long}, and bird feathers.\cite{mueller2023structure} The Namaqua sandgrouse (\textit{Pterocles namaqua}) lives in desert areas, where the male uses its helical-shaped belly feathers to collect and transport water nearly 20 km with a speed of up to 60 km/h \cite{mueller2023structure}. Fibers with helical/spiral shapes have also been shown to effectively collect moisture from the atmosphere, where Vogel \& M\"uller-Doblies \cite{vogel2011desert} found parts of the flora in the semidesert of Namaqualand in South Africa have evolved into a circinate, helical, tortuous, or serpentine shape to increase their water harvesting potential. Despite the broad relevance of liquid transport in soft spring-like systems, these flows have received little attention. Below, we show what happens when the spring is very soft and affected by capillarity, i.e., a fairly small elastocapillary number, which leads to complex static and dynamic interface configurations as well as a range of flow regimes.  

A droplet deposited on a single fiber can adopt both a barrel and a clamshell shape.\cite{gilet2010droplets, mchale2001shape} If a droplet instead is deposited on multiple connected fibers that are parallel or with a helical twist, \cite{leonard2023droplets, kern2024twisted, van2024droplet} the energy landscape becomes more complex with bi-stable regions. Bostwick and Steen \cite{bostwick2015stability} described the stability of the liquid bridge between coils of a helical spring with an accompanying diagram corresponding to the pressure and volume of the liquid bridge. Two equilibrium shapes of a droplet were found in a non-deformable spring, where the stability limit between the two states was shown experimentally and analytically by Lowry and Thiessen. \cite{lowry2007fixed, may2008microgravity} The liquid-air interface in the helical spring is named as a helicosysmmetric interface if it is symmetric about the axis. This stability limit suggests that the pitch-to-spring radius ratio should be $<\pi/\sqrt{3}$, to sustain the helicosymmetric shape inside the helical geometry. However, these equilibrium configurations and stability criteria are only explored when the helical spring does not deform. 

Helical spirals made by stimuli-responsive materials, e.g., nylon,\cite{haines2014artificial, haines2016new, leng2021recent} silk,\cite{lin2020ultrastrong, jia2019moisture} siloxanes,\cite{zhao2023hydrogel, zhou2024engineering} hydrogels, \cite{gao2021chemical, wu2024spider} metal-polymer composite, \cite{li2012superelastic} and liquid crystals,\cite{ryabchun2023light, yang2025active} and have been used to mimic muscle fibers, sensor applications, and for adaptive or responsive systems. Artificial muscles made up of thermo-responsive nylon fiber can produce impressively 64 times the specific work of a mammalian skeletal muscle and lift more than a hundred times the weight of a natural soleus muscle with the same length and mass \cite{haines2014artificial}. Similarly, helical springs made up of a soft material such as polydimethylsiloxane embedded with magnetic particles have been proposed for strain sensing applications and are also used as flexible optical fibers for biomedical applications \cite{zhao2023hydrogel}. Several studies have demonstrated how external stimuli can contract or expand helical spirals, only possible with surface or bulk modification, as mentioned above. 

The capillary interaction between two coils of a helical spring is similar to two straight fibers \cite{eisner2000defense, bico2004elastocapillary, jung2014capillary, protiere2013wetting, duprat2020controlling} where the liquid meniscus between the fibers pulls them together.
Capillary contraction is also created by the spooling of a soft fiber around the droplets. \cite{elettro2016drop, schulman2017elastocapillary} Ecribellate orb web has a silk fiber when wet, a fiber coil, and a buckle inside the droplet as the capillary force becomes larger than the Euler buckling load. \cite{elettro2016drop} The interplay of the elastic buckling and the capillary forces is also observed in soft helical coils (made of vinyl-polysiloxane) inspired by spider-webs, which have shown an elastocapillary instability when pulled out of a silicone oil bath for a small elastocapillary number. \cite{jung2014capillary, elettro2016drop}. 
Transport of droplets along fibers is relevant for a broad range of applications, which has inspired studies of effects from; fiber geometry, \cite{tian2011controlling, kern2024twisted, lee2022multiple, zheng2010directional, zhu2021asymmetric, vogel2011desert, yang2025active} wind, \cite{cazaubiel2023influence, bintein2019self} coarsening, \cite{li2023dynamics, feng2024short} gravity\cite{may2008microgravity, lowry2007fixed}, single and multiple connected fibers, \cite{duprat2020controlling, moncuquet2022collecting, lowry2007fixed, may2008microgravity,leonard2023droplets, hou2024bioinspired, zhang2023liquid} soft fibers,\cite{van2021dynamics} non-newtonian fluids, \cite{ghannam1997experimental} and so forth. Our study complements existing literature and points to a new route for potential flow control and capillary actuations, as we demonstrate below for the static and dynamic properties of droplets in soft springs.  


\section{Results and discussion}
\subsection{Experimental system}
We develop an experimental system consisting of a soft helical spring of length \textit{L} (initial length of 10 cm, variable with the pitch of the helical spring due to stretching) and radius \textit{R}=0.77 ($\pm0.1$) mm as shown in Figure 1(a). The spring is custom-made of polyester with a Young's modulus ($E$) of 2.5 GPa and fiber radius $r=0.05$ mm, where data for different radii are shown in the supplementary information (SI5). The method to prepare the helical spring is found in SI1 (Materials and methods). The pitch, $\uplambda$, is defined as the length between two coils and is varied by stretching the spring so that it also leads to changes in the spring coefficient (measured independently). The spring is attached to a solid base at both ends to help minimize radial deflections during the droplet sliding experiments. The spring constant $k$ is measured by adding a small weight to its end, which is found to follow Hooke's law, see SI2 for details, where the elastic modulus $E$, the fiber radius $r$, spring pitch $\uplambda$, and spring radius $R$ can all be used to alter $k$. The ratio between the spring constant  $k$ and the surface tension coefficient $\gamma$ helps define an elastocapillary number $N_{\mathrm{EC}}=k/\gamma$\cite{jung2014capillary}, where $N_{\mathrm{EC}}<1$ surface tension is expected to deform the spring (as shown below). The springs are surface treated so that they are nearly perfectly wetted by droplets of water or a water-glycerol mixture, i.e., equilibrium contact angle $\approx 0$. These droplets have a volume $V\in [2-10]\upmu l$. Liquid is being continuously dispensed between the coils of the spring and does not initially have a spherical droplet shape.

 \begin{figure}[htbp]
	\centering
		\includegraphics[width=0.85\textwidth]{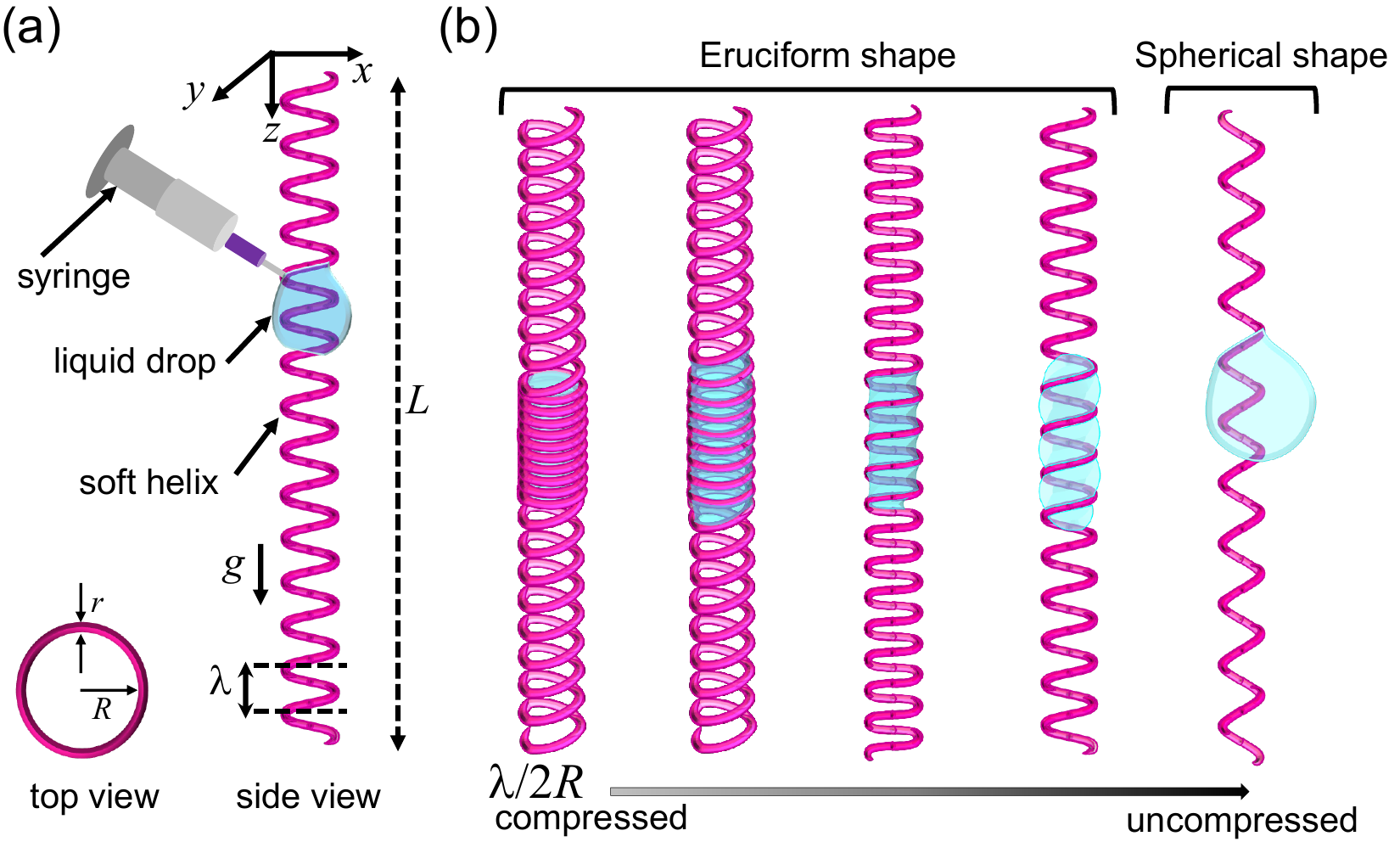}
	\caption{Soft helical spring: (a) schematic of the experimental system (top and side view), with a droplet of volume $V$ and soft helical spring of pitch $\lambda$ and radius $R$, and axis of the helical spring is aligned in z-direction with $g$ gravity and $r$ the fiber radius; (b) from left to right represents the static configurations;  Eruciform (helicosymmetric) to spherical droplet in the on helical spring as increase with $\uplambda/2R$.}
	\label{Figure1}
\end{figure}

\subsection{Static droplet-spring configurations}
We start by systematically examining how the ratio of $\lambda/2R$ versus the droplet volume $V$ affects the resulting interface shape, when the Bond number (ratio between gravity and capillary forces) is still below a critical value, and droplets find a static configuration in the soft spring. $R$ is fixed in our experiments, which means we are altering $V$ for a range of $\lambda/2R$ by slowly injecting liquid between the helical coils. The spring material is nearly perfectly wetting, so the liquid subsequently starts to wet the helical structure. Several different equilibrium interface shapes are found experimentally, as illustrated in the schematic in Figure 1(b), with different degree of compression of the spring.  

We start with a small $\uplambda/2R \leq 0.1$, which we define as regime I. As water meets the spring, a capillary bridge forms between the two coils that spreads across in a radial direction. It forms an ``annulus" with a liquid film filling the gap between the wetted coils, and the spring is still ``hollow" inside, filled with air. A consequence of the liquid meniscus is the generation of a large surface tension force that is sufficiently strong to contract the elastic spring. For small injected volumes ($t\leq 0.06$s), the wetted region forms a capillary bridge extending across multiple coils, where capillarity has contracted the spring so much that the coils nearly touch (Figure 2(a)). By further increasing $V$, the spreading in z-direction halts, likely a consequence of the clamped conditions at spring edges as the dry areas of the spring are stretched, which needs to be compensated by a larger capillary force. Instead, the liquid starts bulging inwards, occupying the air-filled core of the spring while connected to a capillary bridge above and below. The convex curvature inside the fiber continues to grow with $V$, first near the needle tip before being distributed along the radial direction, where the bulge finally seals at the center of the spring, and the middle part of the wetted spring has a liquid-filled core, see Figure 2(a) (see supplementary movie SM1). The interfacial configuration in the spring then consists of a combination of a filled region and hollow regions with a liquid film connecting the coils. Additional experiments have been performed to ensure that the snapshots illustrated in Figure 2 are indeed equilibrated shapes, limited by the time scale for the evaporation of the liquid, see SI3 and SI4. 

If increasing the pitch $\lambda$, i.e., increasing the spacing between coils, leads only to a slightly larger elastocapillary number $N_{EC}$, but the capillary interface takes a very different shape as compared to regime I. In regime II $[0.1<\uplambda/2R< 0.32]$, the liquid immediately fills the interior of the spring, and water bulges from the inside the helical spring between the two coils and forms a semicircular shape between the two fibers (Figure 2(b) schematic) (see supplementary movie SM2, SM3). The droplet-shaped capillary bridge starts spreading between coils with the increase in volume, compressing the spring. For the largest $\uplambda/2R$ in regime II, we also notice that the liquid adopts this bullet-like shape, compressing the spring. Note that, by increasing $\uplambda/2R$, the compression of the spring decreases and the spring constant of the helical spring increases, as shown in the SI2 with a slightly larger elastocapillary number, although $N_{\mathrm{EC}}\ll1$. 

\begin{figure}[htbp]
	\centering
		\includegraphics[width=1.0\textwidth]{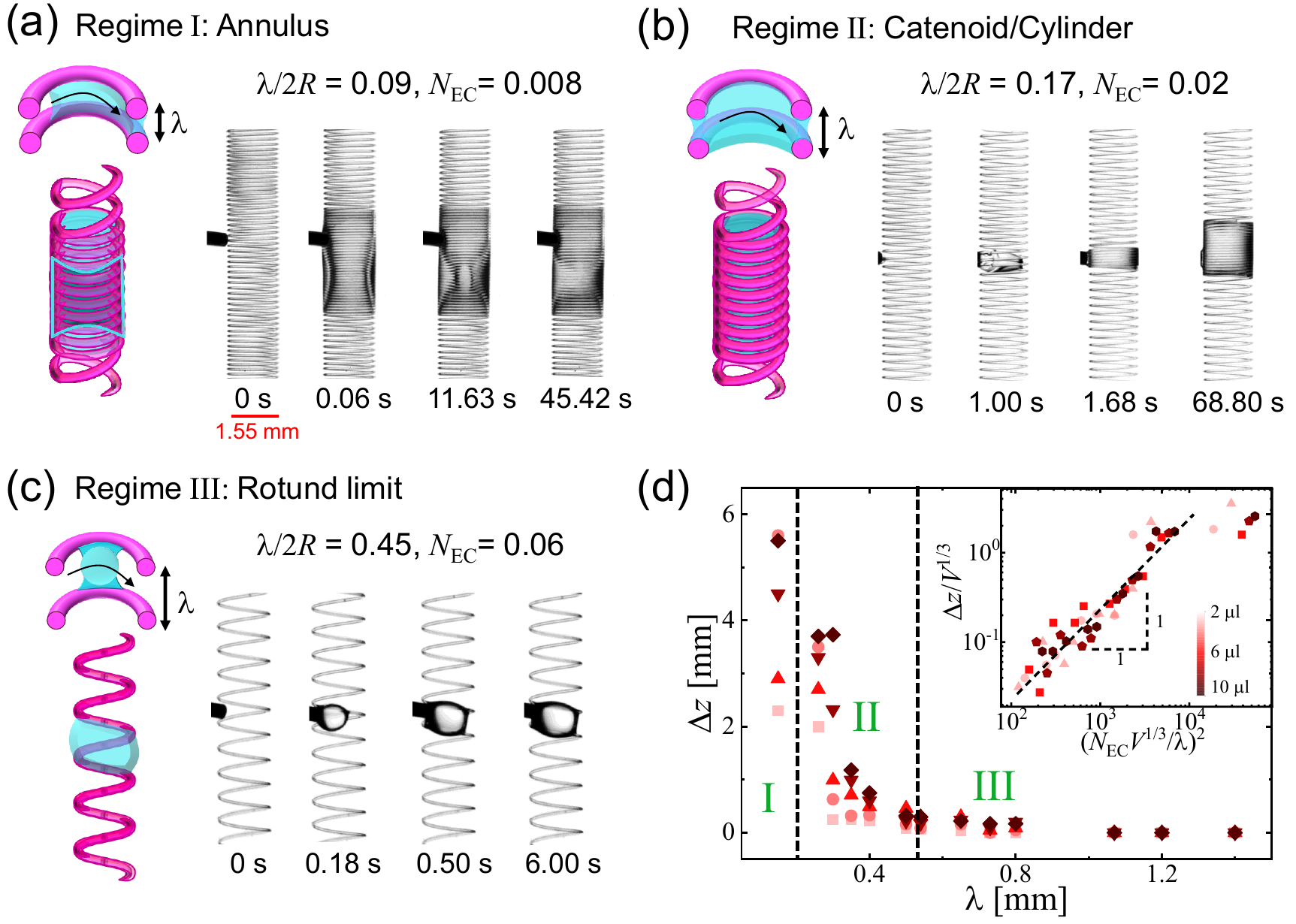}
	\caption{Equilibrium shapes: (a)-(c) optical images at different times from regime I ($N_{\mathrm{EC}}$=0.008) to regime III ($N_{\mathrm{EC}}$=0.06); labeled as annulus, catenoid/cylinder, and rotund, based on their curvature in lateral direction; (d) total compression of the spring in the z-direction ($\Delta z$) with the $\uplambda$, scaled plot of the compression shows dependency on the $\lambda$, $V$, $k$, $\gamma$ (inset) using Equation 1. The scale bar is the same for all the optical images.}
	\label{Figure2}
\end{figure}


If we continue to increase $\uplambda/2R> 0.32$, which we label as regime III, a spherical droplet-shape forms between the spring coils where the interface bulges outwards (see supplementary movie S4). Here, there is nearly no compression of the spring, and the droplet interface is wrapped by the spring. 

To establish the effect of wettability on the compression of the spring, a hydrophobic spring was also prepared, and the optical images are shown in SI5. It is clear that the wettability plays a major role in the compression of the spring because no compression is observed for a spring made up of a hydrophobic fiber. Similarly, to check the effect of the elastic modulus of the material, a copper fiber is used to prepare a solid helical spring (SI6). Since the elastic modulus of copper is $\sim$80 GPa, which is 40 times higher compared to the polyester/nylon fiber, it shows no compression even for a value of 0.09 of $\uplambda/2R$, while the flow regimes are still preserved but different values of $\uplambda/2R$. In all cases, the length of the helix is sufficiently large compared to the deformation of the helix to avoid the error in compression dynamics. Additional experiments have also been conducted with a helical spring composed of 0.2 mm fiber diameter, having an average spring constant of 0.06 Nm$^{-1}$, which exhibits very small capillary-induced deformations (see SI7). So, the compression of the spring depends on the balance between the capillary and elastic forces between the coils.

In the static spring configurations, we measure directly the contractility $\Delta z$ (see SI8) of the spring by comparing the undeformed spring with the final placements of the wetted coils. The capillary force can be estimated based on the $n$ number of wetted coils, where the capillary pressure $\sim \gamma/\lambda$ giving a scaling relation for the net force induced by capillarity to be $F_{\mathrm{cap}}\sim n R^2\gamma/\lambda$, where $n=(V/\pi\lambda R^2)$ is related to the injected liquid volume $V$ and the spring radius $R$. The capillary force is balanced by a the elastic force exerted by the spring which we assume follows Hooke's law $F_\mathrm{E}= k \Delta z$, with $\Delta z$ the compression of the spring in the z-direction. By balancing these two forces, we get
\begin{equation}
\frac{\Delta z}{V^{1/3}}\sim \frac{\gamma V^{2/3}}{k \uplambda^2}\sim \left (N_{\mathrm{EC}}\frac{V^{\frac{1}{3}}}{\lambda}\right)^2.
\end{equation}
The scaling relation gives a prediction for the potential capillary contractility of the spring as a function of the pitch $\uplambda$, droplet volume $V$, and the effective spring coefficient $\kappa$ that is linked to the material choice and spring geometry. To test the scaling prediction, we extract $\Delta z$ from the static experiments, as shown in Figure 2(d), where we have also indicated the approximate transition regions between the different regimes described above. In these experiments, $\uplambda$, $\kappa$, and $V$ are changed, where we notice a dramatic effect in terms of the induced compression as $\lambda$ is decreased. The dimensional data in Figure 2(d), see also inset, is recast into dimensionless form based on Equation 1, which we see follows the predictions well, given that both hollow (liquid bridge between coils) and filled states can be found for the smallest realisation of $\uplambda/2R$. 

\subsection{Capillary flow in a soft spring}
When the injected volume $V$ exceeds a critical value, the droplet goes from a static shape in the springs to a state when it is flowing downwards. It is gravity $g$ that is here the driving force for the droplet motion, where the interfacial dynamics is a complex interplay between viscous and inertial effects in combination with capillarity and a dynamic wetting (contact line) process. In Figure 3(a) we show a snapshot of the flowing droplets when Reynolds number $Re$, which the ratio of inertia and viscous forces is $Re\sim\rho v_z R/\upmu\in[1-40]$ for $\lambda/2R\in [0-0.9]$, where $\rho, v_z, \upmu$ are the density, downward velocity, and viscosity of the droplet, respectively. For the snapshot shown in Figure 3(a), the Capillary number, which is the ratio of viscous force and capillary force, is $Ca\sim \upmu v_z/\gamma\in [2\times10^{-5}-3\times10^{-4}]$, as we change $\lambda/2R\in [0-0.9]$. The droplet flows in soft springs are contrasted with the flow of a droplet in a perfectly wetting capillary tube, i.e., similar to $\lambda/2R=0$. Figure 3(a) clearly illustrates that the soft spring has a highly non-trivial effect on the flow as compared to the capillary tube. For a perfectly wetting capillary tube, we notice the characteristic concave shape of the interface with a wetting-like front ahead of the droplet as it moves downwards along the z-direction. Even with a fairly small $\uplambda/2R=0.16$, the concave shape has disappeared, and the interface at the front and back is nearly flat (see supplementary movie SM5). As we further increase the pitch $\uplambda/2R\approx [0.3-0.8]$, the front and back of the droplet form instead a convex shape where it looks as if it is more and more ``squeezed" in the spring as the pitch increases (see supplementary movie SM6). Here, the droplet adopts an Eruciform shape much similar to a caterpillar, and counterintuitive as one might have anticipated such a shape for a hydrophobic material in contrast to the nearly perfectly wetting springs in Figure 3(a). 

As $\uplambda/2R\geq0.8$, a nearly spherical droplet is formed in the spring (see supplementary movie SM7). These different shapes provide a clue that there are also very different flows within these droplets. In particular, we notice that for small $\uplambda/2R\approx0.16$ the apparent contact line is the same as the advancing front of the droplet that advances as it wets the spring, consequently compressing the spring at the droplets leading edge while spring coils relax as they ``leave" the droplet's trailing edge (Figure 3(a)). While for the Eruciform droplet interface $\uplambda/2R>0.2$ appears to move ahead of the apparent wetting contact line. For a spherical droplet, its leading edge appears to follow the apparent spreading front. The dashed lines in Figure 3(a), second row, show the droplet front/single advancing front, while the blue and red dashed line shows the difference in advancing front and contact line.

To quantify in more detail the differences between these flows, we extract from the experiments the mean velocity in z-direction $v_z$ and the azimuthal velocity $v_{\mathrm{\phi}}= s \omega$, with $s$ the radial distance from center of helical spring axis and $\omega$ the angular velocity based on the droplet's center of mass (see SI9). By extracting these velocities, we show that there is a non-monotonic relationship between $v_z$ and $\uplambda$. The non-monotonic behaviour of $v_z$ is somewhat counterintuitive as the spring's wetted area is larger when the droplet is in Eruciform versus a spherical shape. Moreover, there appears to also be a range of $\uplambda$ where the droplet may adopt both an Eruciform and a spherical shape, which is a bi-stable region and illustrated by the shaded areas in Figure 3(b). By comparing data from these two regimes/droplet shapes, it is clear that the droplet with an Eruciform shape moves downward much faster compared to the corresponding spherical shape, as shown in Figure 4, as also seen within the bistable regime (Figure 3(b)). It is only the spherical-shaped droplets that have a $v_{\mathrm{\phi}}\neq 0$, as the Eruciform-shaped droplet has a center of mass lying on the axis of the spring, i.e., $v_{\mathrm{\phi}}=0$. The rotation velocity $v_{\mathrm{\phi}}$ of the spherical droplets around the helix is plotted in Figure 3(c). The larger the pitch and the liquid volume, the larger the rotational velocity of the droplet. For a larger pitch $\uplambda/2R>0.65$, and for larger liquid volume, the spherical shape water droplet starts to follow the fiber geometry, and due to that, it gives oscillations in the spring's radial direction. Although the Eruciform-shaped droplets have $v_{\mathrm{\phi}}=0$, it is clear that it has an azimuthal velocity within the fluid as the videos clearly illustrate how the interface appears to be ``arching" through the spring.

\begin{figure}[htbp]
	\centering
		\includegraphics[width=1.0\textwidth]{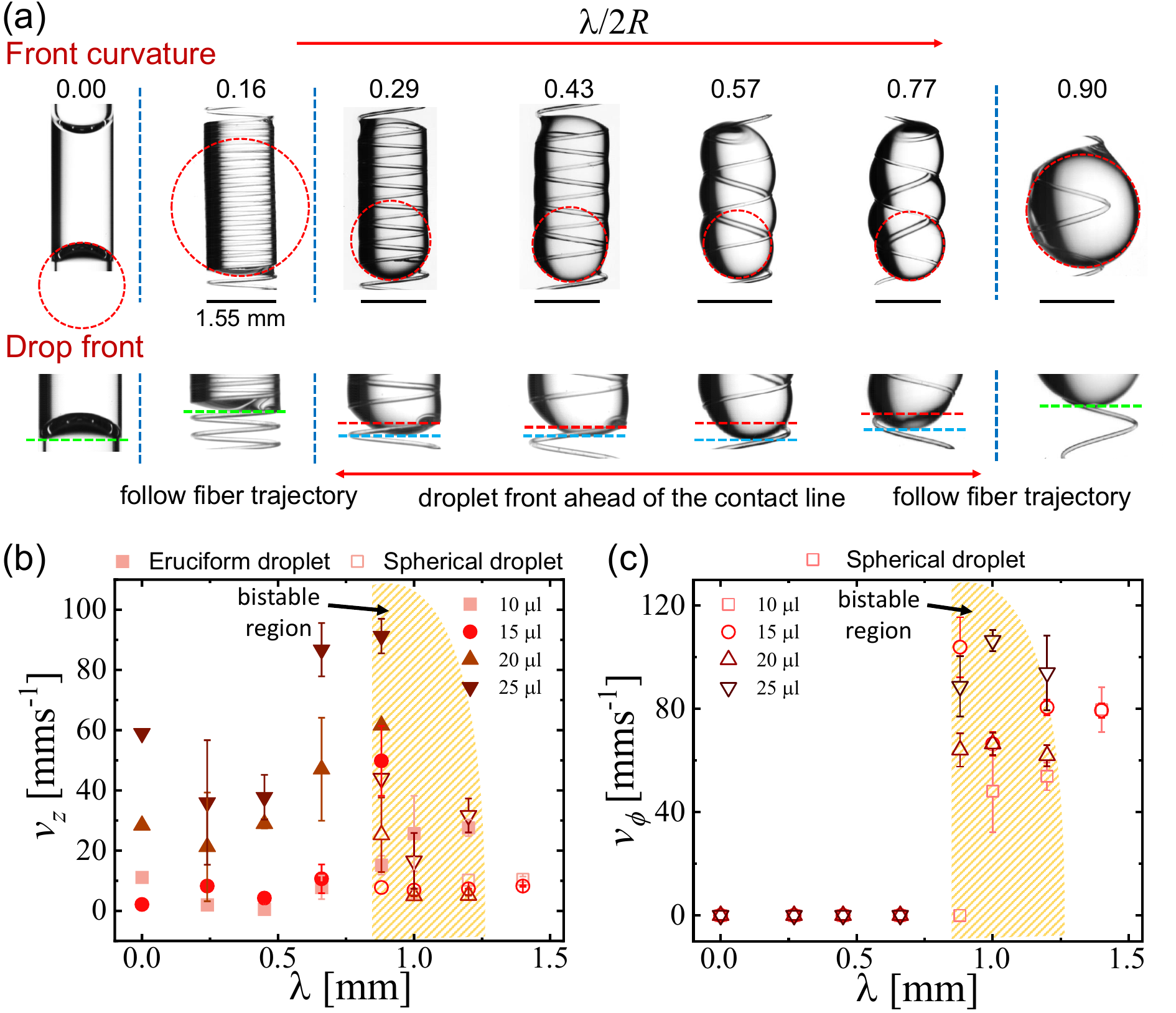}
	\caption{Droplet on a fiber; (a) Eruciform-to-spherical droplet transition with $\uplambda/2R$ for $V=$10 $\upmu$l, where $\lambda/2R=0$ is for capillary tube. The first row corresponds to the droplet equilibrium shapes on different $\lambda/2R$ and curvature (red circle) of the droplet front while moving downward, and the second row shows the droplet front ahead of the contact line with $\lambda/2R$. The green dashed line indicates that the droplet front and contact line positions are the same, while the blue dashed line represents the droplet front and the red dashed line represents the contact line position, if they are different. The scale bar is the same for all the optical images. (b) Downward velocity $v_z$ for an Eruciform and spherical droplet with $\uplambda$, (c) azimuthal velocity $v_{\mathrm{\phi}}$ for an Eruciform and spherical shape droplet. The yellow shaded region is the bistable region.}
	\label{Figure3}
\end{figure}

Why the Eruciform-shaped droplets have a higher velocity as compared to a spherical-shaped droplet can be explained from their internal flow fields as they traverse the helical spring. These flows are illustrated by seeding the liquid with 0.2$\%$ of 10 $\upmu$m silver-coated hollow glass spheres. The analysis of the downward flow is done by using the particle image velocimetry (PIVlab) tool in MATLAB, and averaged over multiple time sequences as shown in Figure 4(b) for an Eruciform and spherical droplet.
The mean internal velocity in Cartesian coordinates, $\overline{v}=(v_z^2+v_x^2)^{1/2}$, where $v_x$ is the velocity in the x-direction, is shown for an Eruciform and spherical droplet shapes in Figure 4(b) for $\uplambda/2R=0.65$ and $V=10\;\upmu$l. The heat map shows that for an Eruciform droplet, internal flow is not only downward, but it also has some rotational flow while moving downward. The high internal velocity (red color in Figure 4(b)) appears when the droplet follows the fiber curvature and takes turns to move downward. While for a spherical droplet, the internal flow is also rotational, but the angle of this rotational flow with gravity is low compared to an Eruciform droplet. This internal rotational flow appears with the increase in the pitch; we have observed no internal flow for a small pitch. With the $\uplambda/2R>0.43$, the droplet interface in the lateral direction becomes convex, and with a more complex internal flow having $Re>1$ (see SI10). While an Eruciform droplet is moving downward, it experiences the spiral curvature; due to this, the droplet has a different flow orientation at the advancing and receding ends. The flow in the liquid makes an angle with respect to the direction of gravity, so that the advancing front increases the outward flow, which enhances the overall downward flow. We have considered that the center of the mass of the droplet is aligned with the axis of the helical spring, resulting in zero azimuthal velocity around the spring. For $\uplambda/2R>0.43$, when the droplet is in spherical shape, internal flow as well as the droplet rotates around the axis of the helical spring (low internal vorticity) with a small angle with the gravity compared to an Eruciform droplet, resulting in an increase in azimuthal velocity and reducing $v_z$ as shown in Figure 3.

\begin{figure}[htbp]
	\centering
		\includegraphics[width=1.05\textwidth]{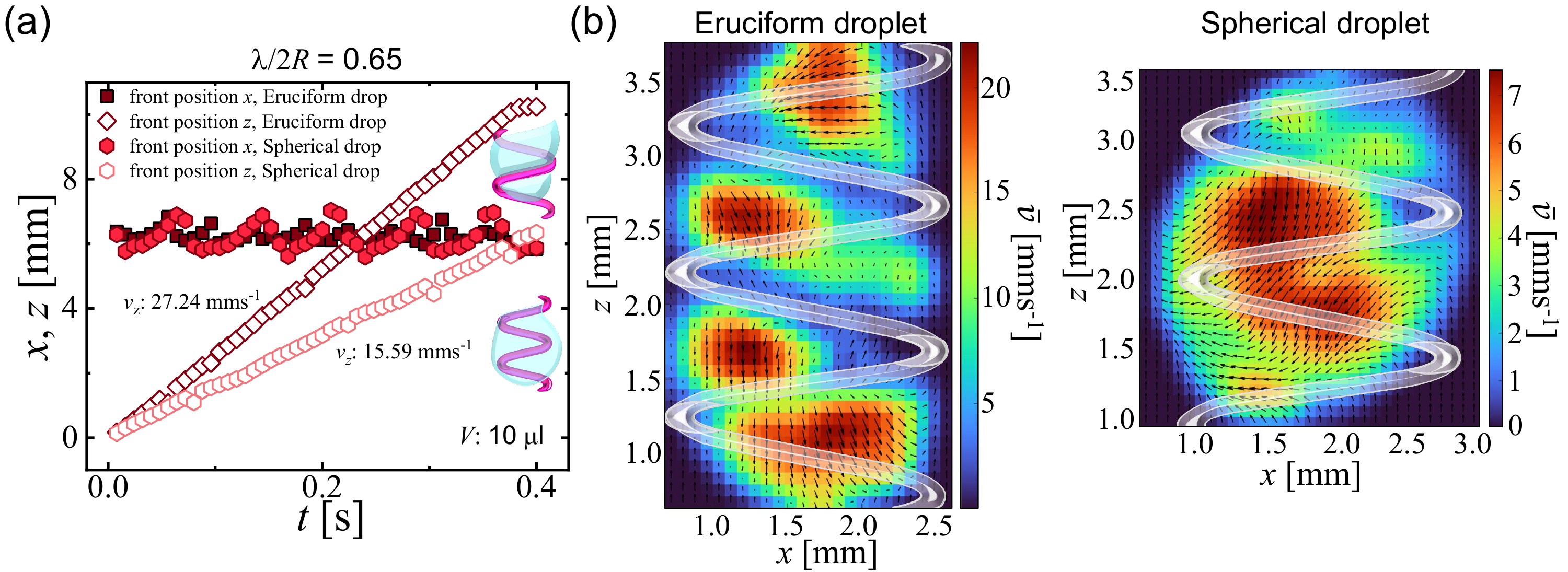}
	\caption{Flow inside helical spring; (a) x,z-position of the advancing front of an Eruciform and spherical droplet for $\uplambda/2R=0.65$ and $V$=10 $\upmu$l, (b) PIV images for internal velocity $\overline{v}=(v_z^2+v_x^2)^{1/2}$ for an Eruciform and spherical shape droplets for $V$=10 $\upmu$l, and $\uplambda/2R=0.65$. A helical spring schematic is added over the PIV images to guide the droplet shape.}
	\label{Figure4}
\end{figure}

The larger the value of $\uplambda/2R$, the less the Eruciform droplet is influenced by the presence of the fiber geometry, while the spherical-shaped droplet starts experiencing the wall adhesion (droplet is pinned between the coils while moving downward) and behaves as a drop behaves in the Wenzel state (see SI11). 
Since the droplet is moving in the helical spring and following the spring geometry, for $\uplambda/2R>0.9$ and volume $>$20 $\upmu$l, due to centrifugal force, droplets start detaching from the helical spring. Also, we observed that for $V>$ 30 $\upmu$l,  small liquid droplets were started ejecting from the receding part of the Eruciform droplet, see SI12.

\subsection{Influence of viscosity}
\begin{figure}[htbp]
	\centering
		\includegraphics[width=1.0\textwidth]{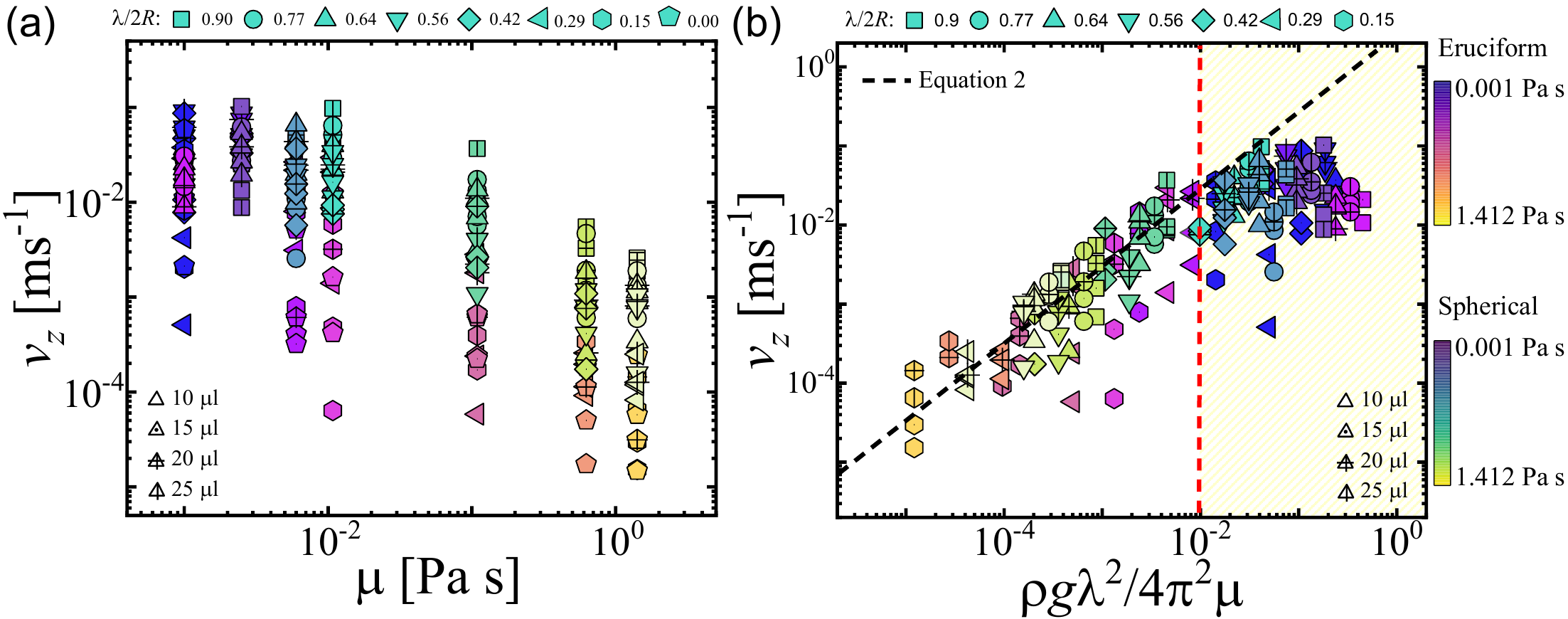}
	\caption{Droplet flow dynamics; (a) $v_z$ of the Eruciform and spherical droplet for $\upmu\in[0.001-1.412]$ Pa$\cdot$s, (b) scaled $v_z$ of the Eruciform and spherical shaped droplets with $\upmu\in[0.001-1.412]$ Pa$\cdot$s, $\lambda/2R\in[0.15-0.9]$ and $V\in[10-25]\upmu l$. The black dashed line is the scaling prediction in Equation 2 (slope 1), and the vertical red dashed line indicates the transition where $Re$=1, where to the right of the line $Re>1$.}
	\label{Figure5}
\end{figure}
We also test the effects of the viscosity $\upmu$, which is varied by altering the fraction between water and glycerol, with $\upmu\in [0.001-1.4]$ Pa$\cdot$s (SI13). Viscosity can alter the flow dynamics, but it turns out to also affect the Eruciform-to-spherical shape transition, likely through changes in the capillary number. For water and water-glycerol solutions of $\upmu$ up to 0.006 Pa$\cdot$s, the transition is observed around $\uplambda/2R=0.43$, but the transition is observed at $\uplambda/2R=0.29$ for larger $\upmu$. This shape transition involves the time scale associated with the dispensing rate to flow inside the helical spring. For large $\upmu$, the liquid adopts a spherical shape first and then starts flowing along a helical fiber. Similarly, if enough time is not provided to spread the liquid between the coils, the liquid always comes up as a spherical droplet shape in the spring. So, the boundary between different regimes depends on $\uplambda$ and the dispensing rate of the liquid when the viscosity is high. $v_z$ of the droplets in an Eruciform droplet and a spherical droplet is shown in Figures 5(a), respectively. Intuitively, the increase in viscosity decreases $v_z$. 


To rationalize the downward speed $v_z$ of the droplets, we turn to a scaling prediction by balancing $F\times v_z\sim \upphi_\upmu$ where $F=\rho V g$ is the gravitational force and $\upphi_\upmu$ is the viscous dissipation. If using the center of the spring as a point of reference and adopting cylindrical polar coordinate system ($r^*$, $\phi$, z) we can estimate the viscous dissipation in the bulk of the droplet to be $\upphi_\upmu=\upmu \int \left( \frac{\partial v_z}{\partial r^*} \right)^2 \, dV\;+\;\upmu\int \left( r^*\frac{\partial}{\partial r^*} \left(\frac{v_\phi}{r^*}\right) \right)^2 \, dV$. The droplet has both a vertical and rotational flow within, where the latter is believe to dominate given that parts of the droplet is exposed to air with a free slip. By using this assumption the main part of the dissipation would come from $\upphi_\upmu\approx \upmu\int \left( r^*\frac{\partial}{\partial r^*} \left(\frac{v_\phi}{r^*}\right) \right)^2 \, dV$, where $r^*\sim R$ (the spring radius). 
Inserting these scaling relations together with $v_\phi=2\pi R v_z/\uplambda$, connecting the droplet flow $v_z$ and the spring geometry with $v_\phi$ gives,
\begin{equation}
v_z\sim\frac{\rho g\uplambda^2}{4\pi^2\upmu}
\end{equation}
The scaling prediction gives a link between the rotational flow within the droplet and the flow speed through the spring. We use Equation 2 to rescale the velocity for a range of viscosities $\upmu$, volumes $V$ and pitch $\uplambda$ in Figure 5(b). The scaling prediction is expected to hold when $Re<1$, and, as expected, the data deviate from the scaling prediction for $Re>1$ as shown in Figure 5(b). When $Re>1$, the inertia starts to dominate over the viscous effect in the flow. There are other factors also involved, such as enhanced internal rotational flow and secondary flow, spring oscillation, wall adhesion, and the droplet is pinned between two spiral coils with similarities to a Wenzel state.\cite{aussillous2000quick}

\subsection{Lifting weights with droplets and actively tuning droplet flow}
Now we turn to illustrate how these understandings of equilibrium and droplet dynamics can be used to demonstrate actuation and active flow control. We start by illustrating how actuation is acheived by converting capillary energy into spring energy. To demonstrate this, we attach a weight $m=[39, 49] \pm1$ mg to the end of the soft spring as shown in Figure 6(a). After the weight has been placed, the spring is allowed to relax before liquid is dispensed. For a $m=39$ mg weight and $m=49$ mg, the stretched spring has an $\uplambda/2R=0.12$ and 0.16, respectively, and water from [2 -20] $\upmu$l  (weight $\approx 2-20$ mg) is dispensed on the spring. As water is dispensed, the spring contracts and, after an equilibration, reaches a static shape. The total deformation of the helical spring for the two added weights is shown in Figure 6(b). This total compression corresponds to the injected liquid volume, which has its potential surface force. To estimate the potential work by tensile actuation to do work, generated by the surface forces, we extract $\Delta L/L\times100$, with $\Delta L$ the spring deformation and $L$ the spring length. We note that only a small fraction of the coils are wetted and as such are responsible for the compression. If additional coils have been wetted, one may expect a total deformation could be much greater. As such, the specific work measured here corresponds only to the wetted area of the spring; optical images are shown in Figure 6(a). The highest tensile actuation observed is 73$\%$ for a weight of $m=0.049$ g, but if the total length of the spring is taken into account, it would reduce to 12 $\%$. The maximum specific work ($F_{\mathrm{cap}}\Delta L/m$, where, m is the attached mass) done by the 20 $\upmu$l droplet is 54 J/kg, which is higher than the specific work done by the mammalian skeletal muscle, 38.6 J/kg, but less than the hydrothermal-induced compression on coiled nylon, 2860 J/kg studied before\cite{haines2014artificial}, i.e, only 10$\%$ of the total length is wetted, and with a small volume. It is noted that the system here is not optimised for the amount of wetted area of the spring. The specific work is calculated when only 10$\%$ (length of the droplet/total length of the spring) of the length of the spring is responsible for lifting the weight. 

\begin{figure}[htbp]
	\centering
		\includegraphics[width=0.9\textwidth]{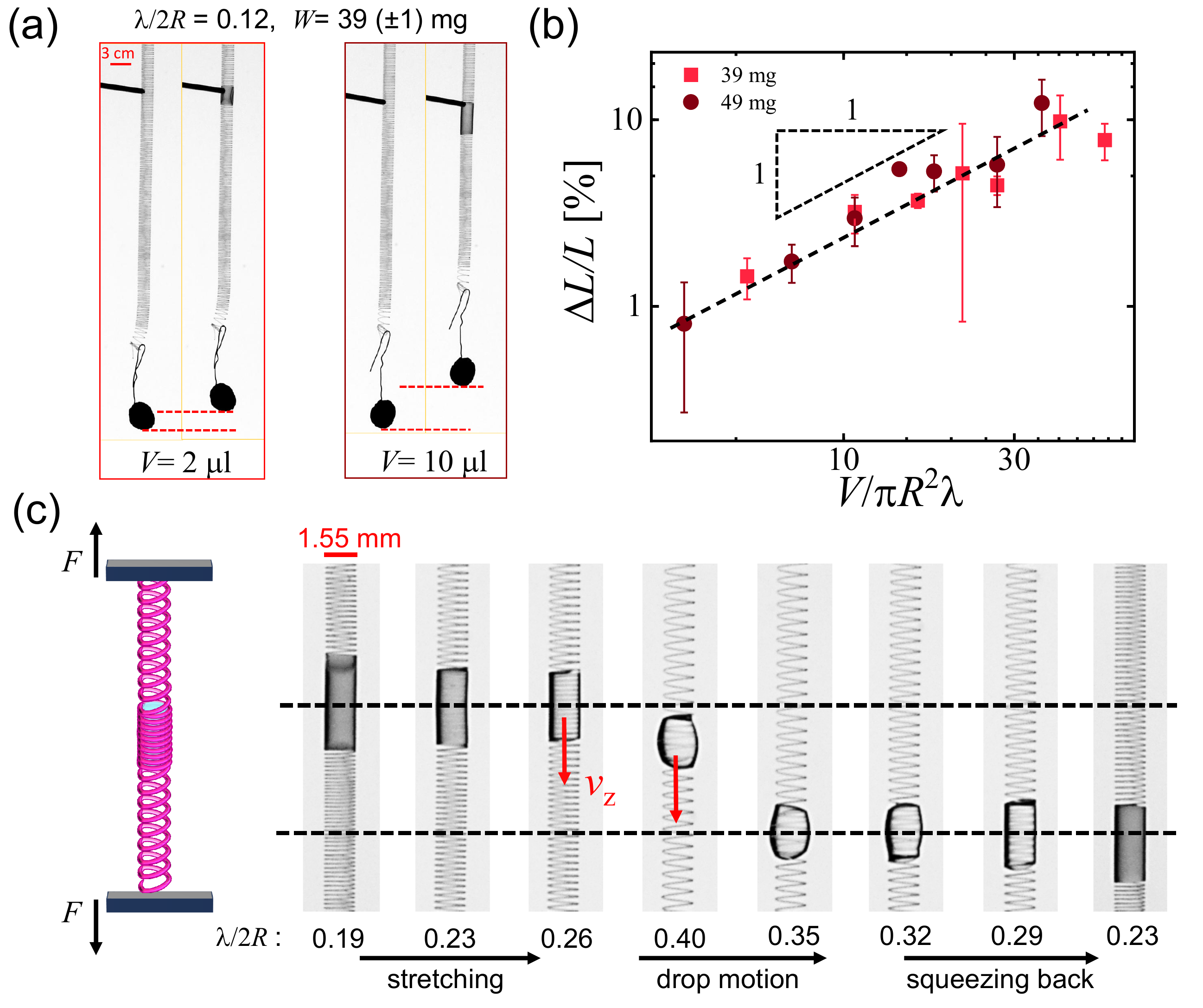}
	\caption{Lifting weights with droplets; (a) optical images of the spring deformation weight due to capillary force. (b) tensile actuation $\Delta L/L [\%]$ for different droplet volumes with the weights $m=0.039$ ($\pm0.01$) g and $m=0.049$ ($\pm0.01$) g, (c) droplet flow control by active stretching and squeezing the spring while the spring is filled with 10 $\upmu$l water.}
	\label{Figure6}
\end{figure} 

Next, we apply these concepts and knowledge to control droplet dynamics in a soft spring, demonstrating this as a proof of concept for flow control. The analysis of static shapes illustrates how droplets can ``freeze" into different configurations, while the droplet flow analysis indicates that the spring wavelength $\lambda$ can control the droplet flow regime. We utilize this method as an active mean to both stop and induce droplet flow in these springs, as illustrated in Figure 6(c). Water is deposited in the spring with $\uplambda/2R=0.19$ and finds a static form in the spring (regime I). The wavelength is then changed by stretching the two ends, and as $\uplambda/2R=0.26$ we notice that the droplet flows with an Eruciform shape in the spring. Imagine that we would now also want to control the speed of the droplet, where we can utilize the insights from the different flow regimes as a function of $\uplambda/2R$. As the spring is further stretched, water transitions from the Eruciform to a spherical droplet shape, significantly reducing the descent speed. Now, this dynamics is reversible, whereupon compressing the spring, the droplet speed picks up, and it forms an Eruciform shape before returning to its static, arrested shape, as seen in Figure 6(c).



\subsection{Conclusion}
We have shown how a soft spring system behaves when introduced to a liquid, when capillarity can significantly compress the spring, i.e., a small elastocapillary number. The spring geometry plays a crucial role in determining both the equilibrium shape and the dynamic flow of droplets. A range of complex interface shapes is determined for the equilibrium shapes of the spring and droplets. Beyond a critical volume, these droplets start to flow, and $\uplambda$ is a key parameter to determine the flow regime. For small $\uplambda$ droplets, flow with cylindrical shape and a velocity profile along the z-direction of the spring. When $\uplambda/2R<0.3$, it takes the shape of a caterpillar, with an Eruciform shape arching through the spring with an increased velocity and also a slight flow in the azimuthal direction, although its centre of mass stays put at the centre of the spring. When increasing further $\lambda/2R>0.77$, the droplet adopts a spherical shape, significantly reducing the settling velocity and with a motion of its centre of mass in the azimuthal direction.  

Through an understanding of the equilibrium and dynamics of droplets in soft springs when the elastocapillary number is small, we showed how these effects can be adopted to induce actuation and flow control. There are many more interesting aspects in these capillary slinkies that warrant future studies, which this work can hopefully stimulate, including the decoupling of gravitational influences, liquid film deposition, partial wetting, as well as computations of the equilibrium shapes and the droplet flow. 


\begin{acknowledgement}
We would like to thank Dr. Susanne Liese for stimulating discussions. The authors acknowledge funding from the Research Council of Norway (project number: 355133), the UiO Sustainability Initiative, and an ERC grant (TWIST, grant agreement:101169717). Funded by the European Union. Views and opinions expressed are however those of the author(s) only and do not necessarily reflect those of the European Union or the European Research Council Executive Agency. Neither the European Union nor the granting authority can be held responsible for them.

\end{acknowledgement}
\bibliography{acs-achemso}

\begin{suppinfo}
\noindent\textbf{SI1: Materials and methods}
\par
\noindent\textbf{Materials:}
Polyamide (Nylon) and Polyethylene terephthalate (polyester) threads of diameter 0.1 mm were purchased from a local Haberdasher, Oslo, made by G{\"u}termann (Col.1001), Germany. Florocarbon PVDF fiber was purchased from Seaguar Gold Label 100$\%$ florocarbon Fishing Line leader, 2 lbs, 25 yds - 02GL25. Copper wire was purchased from a local shop in Oslo. Glycerol and acetone were purchased from VWR, Norway, and used without any further purification. Aluminium rods of diameter 1 mm were used as a mandrel, purchased from a local shop in Oslo, Norway. A drilling machine (Cocraft LXC DD18) having high RPM 0-1500 and low RPM 0-400 was used to make helical geometry. Silver-coated hollow glass spheres (S-HGS-10) of 10 $\upmu$m were purchased from DANTEC DYNAMICS, used as a seeding particle for PIV analysis.
\par

\noindent\textbf{Image acquisition and analysis:}
The experimental setup contains the syring pump ($kd$Scientific, Legato@200) to dispense the liquid inside the helical spring at a control rate of 1 $\upmu$ls$^{-1}$. The FASTCAM SA5 Photron high-speed camera was used to record the statics and dynamics of the droplet at 125 fps and 500 fps, respectively. The white light source was used to backlight the system. The recorded sequences taken from the camera are processed using the code written in Python and MATLAB R2023b. The different needle diameters were used from ID 0.26 mm to 2 mm, to dispense different drop volumes. During drop dynamics, the internal flow dynamics were analysed using the MATLAB PIVlab. \cite{thielicke2021particle}
\par

\noindent\textbf{Experimental method:}
Nylon and polyester are thermoplastic materials, and because of their reversible contraction property with temperature, they can be easily molded into any desired shape. \cite{haines2014artificial} Before making a helical spring, the thread of required length was cleaned using acetone in an ultrasonic bath for 10 min. Then thoroughly rinsed with DI water and dried with compressed air. The mandrel of 1 mm diameter was attached to the head of the drilling machine, and the other end was free. The free end of the mandrel was attached with the thread, and a weight of 100 g was hung to the end of the thread to get tension on the thread.  The weighted end of the thread is free to wrap around the mandrel when the drilling machine is rotating at a fixed RPM. The RPM was kept fixed for all the experiments to get the same tension on the thread while preparing the helical spring. Spin the mandrel up to the required helical spring length, and after removing it from the drilling machine, both ends are glued properly so that it remains fixed on the mandrel. To make a helical spring, the wrapped mandrel with thread was heated in the oven at a critical temperature of $\sim$ 160 $^\circ$C for 1 h 30 min in the presence of steam of water to reduce the effect of dryness during preparation. Above this temperature, the material starts oxidizing, and the surface property might get changed. The chosen temperature should be larger than the glass transition temperature of nylon and polyester, which is $\sim$47 $^\circ$C and 60 $^\circ$C, respectively, and less than the melting temperature $\sim$220 $^\circ$C. The critical temperature was chosen by doing multiple experiments at different temperatures and found that only the springs prepared at this temperature remain stable. After taking it out of the oven, the mandrel with the thread is left at room temperature for 1 h to cool down, before removing it from the mandrel. Later, the helical spring is slowly removed from the mandrel, and it will remain in its helical shape. The prepared helical spring was cleaned with acetone and dried. To get a completely wettable state, the spring was oxygen plasma cleaned for 10 min and used in that state. Between every experiment, the helix was cleaned with acetone, followed by plasma cleaning.  For compression experiments, the water drop of different volumes is deposited on a helix. For droplet dynamics experiments, water and water-glycerol solutions were used, having viscosities of 0.001 Pa$\cdot$s to 1.4 Pa$\cdot$s. 
\par

\noindent\textbf{SI2: Spring constant and elastocapillary number:}
\begin{figure}[htbp]
	\centering
		\includegraphics[width=0.8\textwidth]{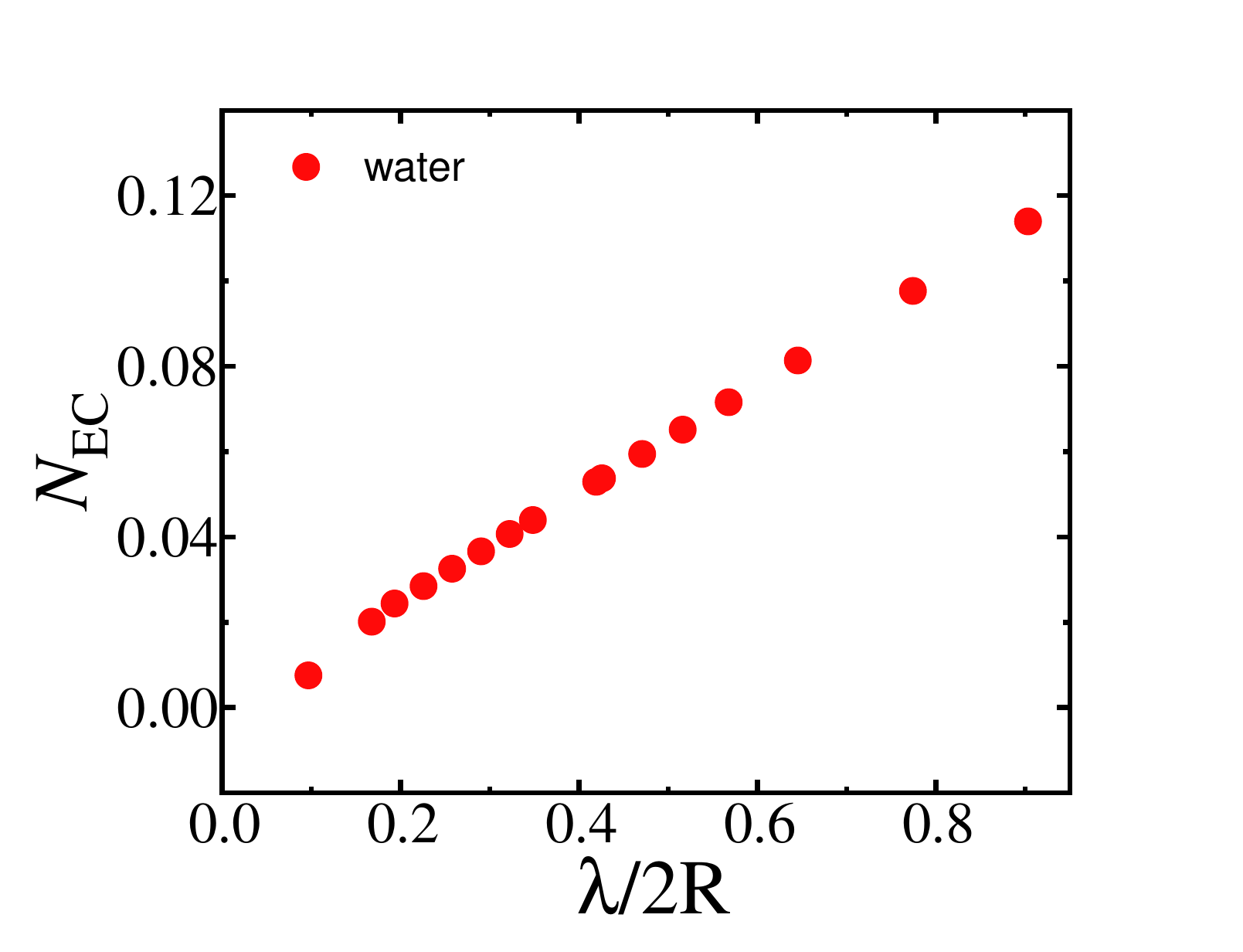}
	\caption{Elastocapillary number with pitch and radius of the helical spring.}
	\label{FigureS2}
\end{figure}
Spring constant for a helical spring 
\begin{equation*}
k=\frac{Ed^4(L-nd_{\mathrm{f}})}{16(1+\nu)(D-d_{\mathrm{f}})^3n}
\end{equation*}
where $E$, $d_{\mathrm{f}}$, $n$, $L$, $\nu$, $D$ are the elastic modulus of the material, diameter of the polyester fiber, number of the active coils, length of the spring, and diameter of the helical spring, respectively.
The spring constant measured experimentally of the helical spring with 1.55 mm diameter and 0.1 mm thread diameter is 0.008 ($\pm$0.002) Nm$^{-1}$ using different weights. The average value of spring constant calculated from the relation is for $\lambda/2R$ =0.09 is 0.0005 Nm$^{-1}$ to 0.008 Nm$^{-1}$ for $\lambda/2R$=0.9, hence $N_{\mathrm{EC}}$ is 0.007 to 0.063. However, $N_{\mathrm{EC}}$ is increasing but $<$1, still the compression is negligible for $\lambda/2R=0.9$. \par

\noindent\textbf{SI3: Compression dynamics for 1 $\upmu$l droplet:}
\begin{figure}[htbp]
	\centering
		\includegraphics[width=0.8\textwidth]{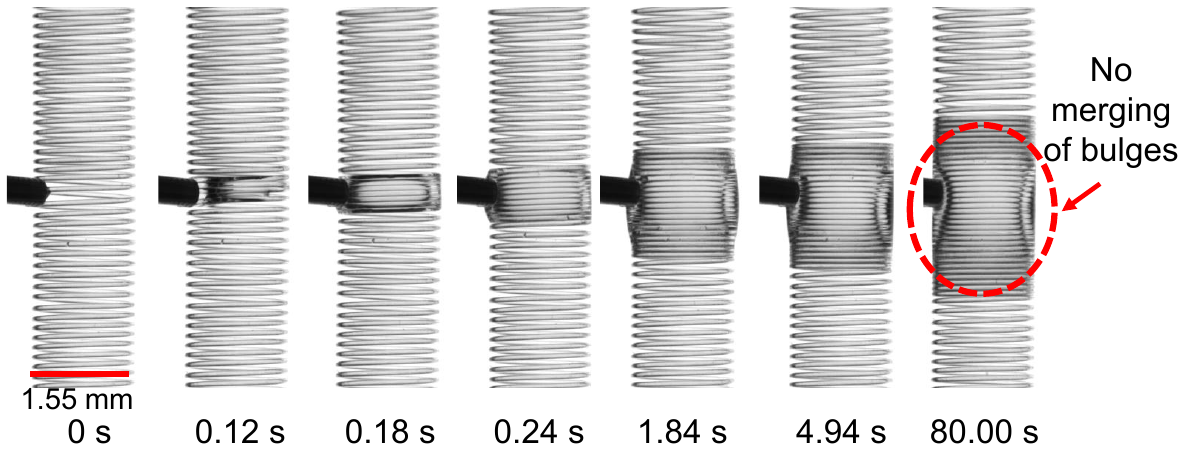}
	\caption{Spring compression dynamics for 1 $\upmu$l water droplet for $\lambda/2R=0.09$. The scale bar is the same for all images.}
	\label{FigureS3}
\end{figure}
For 1 $\upmu$l volume of water, the water spreads between the coils, and after a certain volume, it starts overfilling the fiber spacing. Because the volume dispensed is not sufficient to overfill all the fiber spacings, these overfilled areas have not grown enough to merge with another side bulge. If the system remains in this stage, the overfill liquid spreads back to multiple fibers because the liquid in this curved state is not stable for a long time. No merging observed for water inside the helical cylinder for 1 $\upmu$l volume as shown in Figure S2.\par

\noindent\textbf{SI4: Compression dynamics for 10 $\upmu$l droplet:}
\begin{figure}[htbp]
	\centering
		\includegraphics[width=0.5\textwidth]{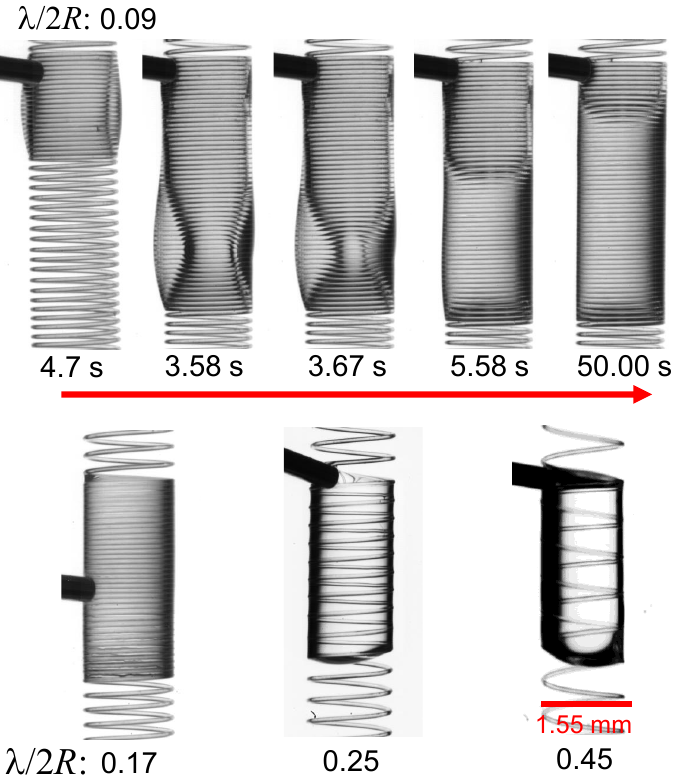}
	\caption{water dynamics for different $\lambda/2R$; first row is for $\lambda/2R$=0.09; second row is for the final configuration of higher values of $\lambda/2R$ =0.17, 0.25, and 0.45. The scale bar is the same for all images.}
	\label{FigureS4}
\end{figure}
For $\lambda/2R$=0.09, the water and fiber interaction is different compared to the other values of the $\lambda/2R$ as shown in Figure S3 (first row). Initially, the liquid spreads between the coils and then starts bulging and eventually merges with the neighboring bulge. After the bulge merges with the other side, continuous dispensing will lead to a filling of the cylindrical structure as shown in Figure S3. For larger volume $>$ 10 $\upmu$l, the weight of the liquid is responsible for the spring sagging in a downward direction, and the liquid column becomes unstable. For higher values $\lambda/2R>$ 0.1, the dynamics is similar to that for corresponding lower volumes, and the final configuration is shown in Figure S3 (second row). 
\newpage
\noindent\textbf{SI5: Compression in hydrophobic fibers:}
\begin{figure}[htbp]
	\centering
		\includegraphics[width=0.6\textwidth]{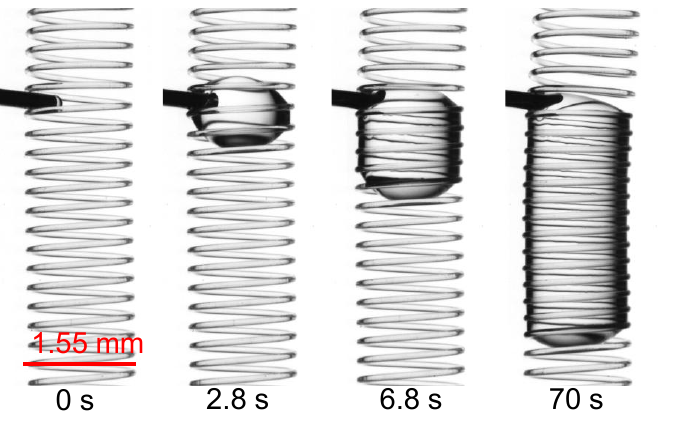}
	\caption{Water on hydrophobic helical spring. The scale bar is the same for all images.}
	\label{FigureS5}
\end{figure}
To check the wettability effect, PVDF fiber (diameter is 0.1 mm, same as hydrophilic fiber used in other experiments) is used to prepare the helical spring, and the measured contact angle on these surfaces is 110$^\circ$. The Spiral coils show slight deformation while dispensing water, but no visible compression is observed for this case, as shown in Figure S4.
\newpage
\noindent\textbf{SI6: Compression of helical spring made up of copper ($E$ = 60 GPa) wire:}
\begin{figure}[htbp]
	\centering
		\includegraphics[width=0.7\textwidth]{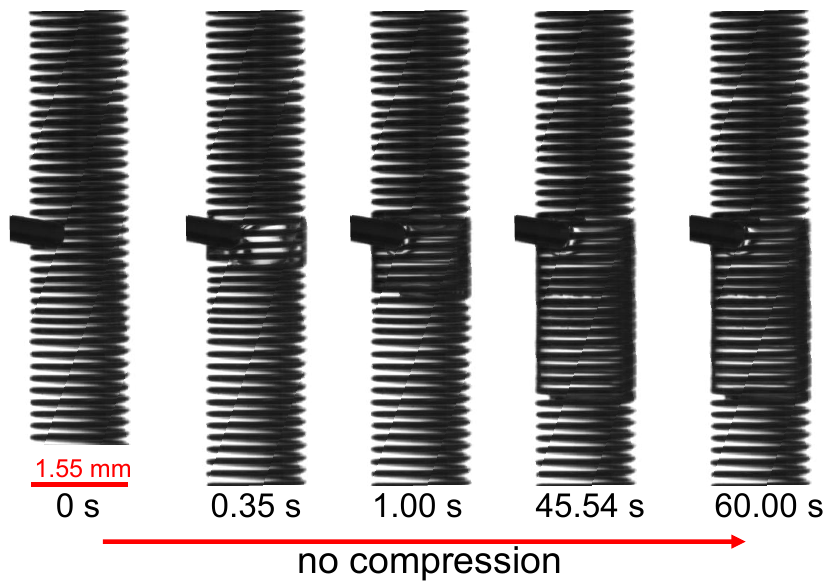}
	\caption{Compression of copper wire due to applied water droplet. The scale bar is the same for all images.}
	\label{FigureS6}
\end{figure}
The helical spring was prepared using 0.1 mm diameter copper wire and hung from two supports at both ends. The 10 $\upmu$l water drop was dispensed and did not find any compression even after waiting long enough, as shown in Figure S5. 
\newpage
\noindent\textbf{SI7: Compression on thicker fiber/evaporation:}
\begin{figure}[htbp]
	\centering
		\includegraphics[width=0.7\textwidth]{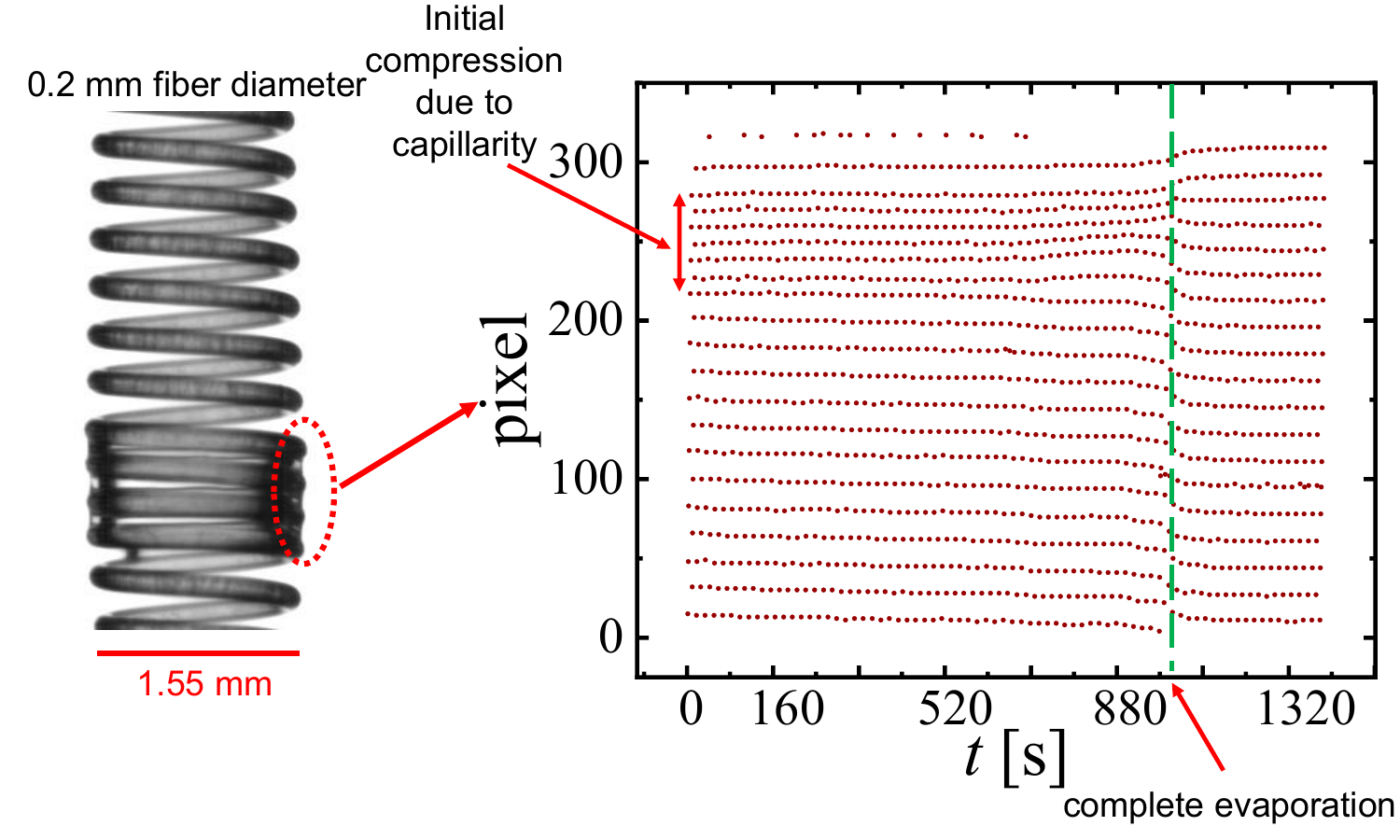}
	\caption{The compression dynamics for a thicker fiber with a 0.2 mm fiber diameter.}
	\label{FigureS7}
\end{figure}
The average spring constant for a helical spring made with 0.2 mm diameter nylon fiber is 0.06 N/m. The elastocapillary number is very close to 1 for this configuration, so it is difficult to see the deformation via capillarity. During the dispensing of the water, because of the larger spring constant, the compression is very small, but after a while, when the liquid starts evaporating, the coils collapse and compress the spring as shown in Figure S6. Then, in the later stage, when the water evaporates completely, the coils regain their original position.\par

\noindent\textbf{SI8: Image analysis for compression:}
\begin{figure}[htbp]
	\centering
		\includegraphics[width=1.0\textwidth]{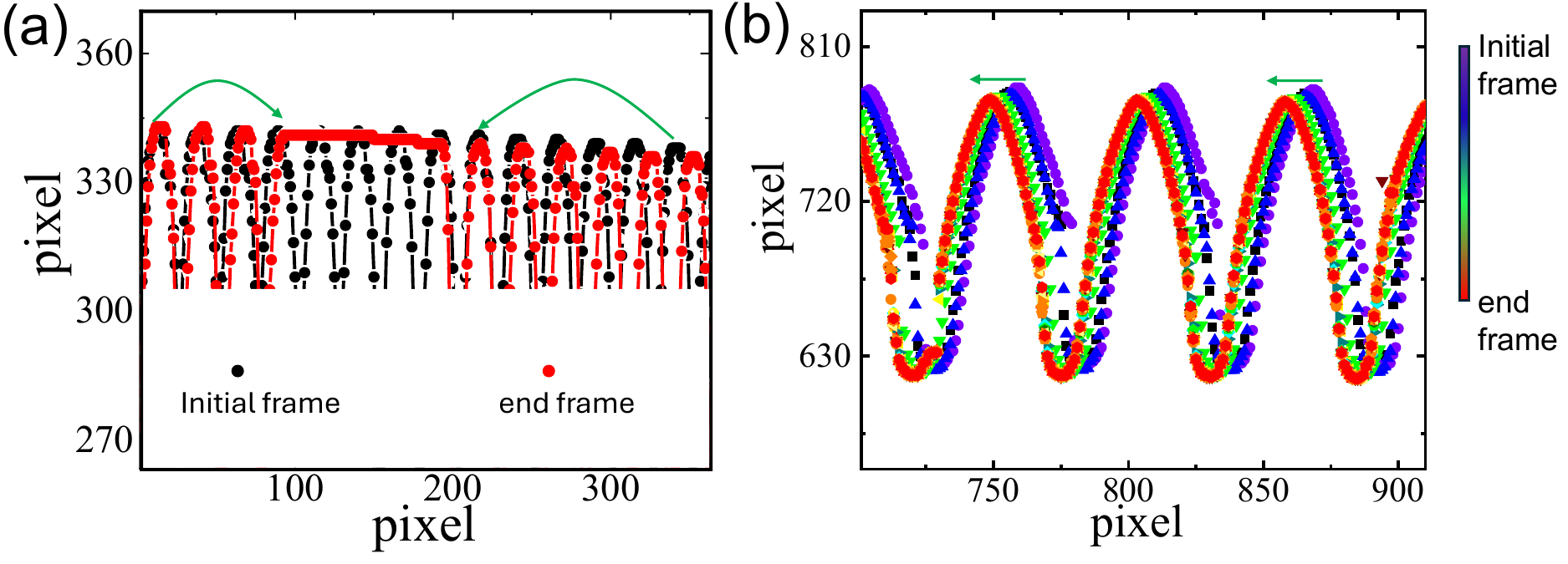}
	\caption{Helical spring dynamics with time; (a) for $\lambda/2R$=0.17, and (b) $\lambda/2R$=0.32.}
	\label{FigureS8}
\end{figure}
The image analysis of the spring dynamics during compression is done by using the code written in Python. During the dispensing of a volume of water, the spring shows some oscillation perpendicular to the axis of the spring; because of this, the image analysis is a little difficult. To measure the compression, first, the edge of the spring is detected using edge detection in Python and overlapped with different frames. Since for $\lambda/2R$= 0.17, the compression is large, only the initial and final frame edges are shown in Figure S7(a). The green arrow indicates the initial to final position of the coil during compression. For larger values of $\lambda$, compression is small, so more frames are shown in Figure S7(b). The peaks in the graph correspond to the edge of the spring coil. \par
 

\noindent\textbf{SI9: Drop dynamics with pitch of the helical spring:}
 \begin{figure}[htbp]
	\centering
		\includegraphics[width=0.9\textwidth]{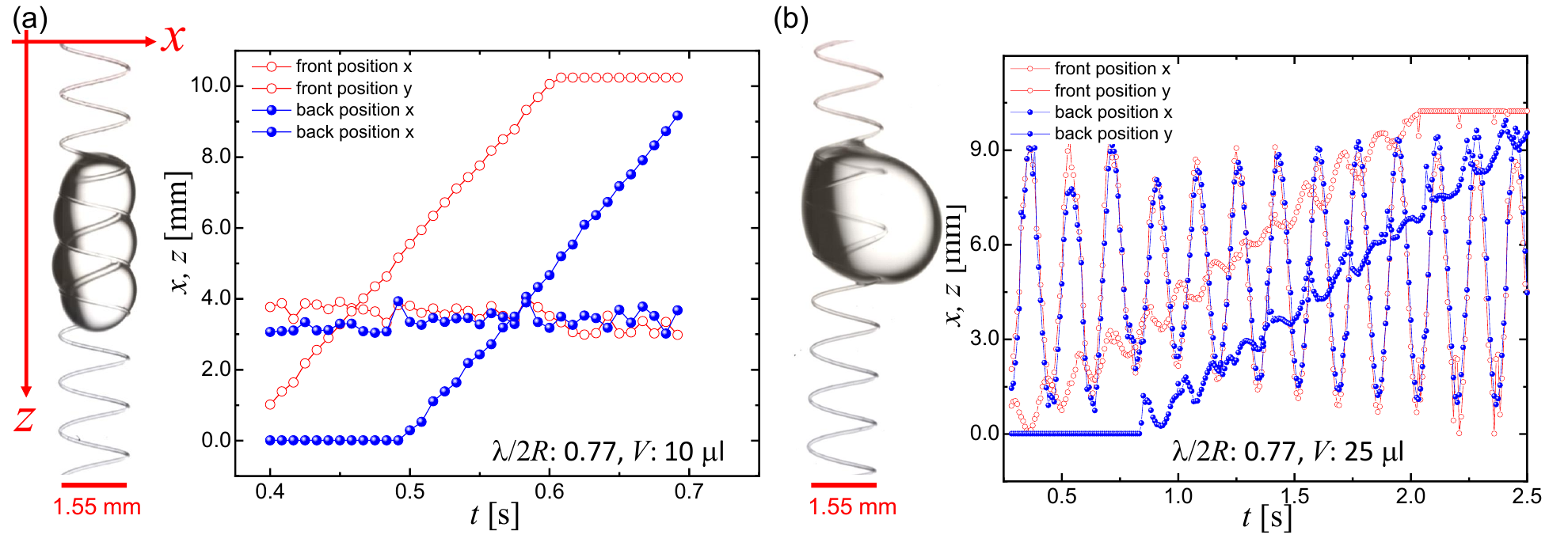}
	\caption{Optical images and x,z-position of an Eruciform and spherical drop of water for $\lambda/2R=0.77$ for 10 $\upmu$l and 25 $\upmu$l.}
	\label{FigureS9}
\end{figure}
Due to high inertia, the spring is also oscillating with the droplets, and centrifugal force acting outward gives the drop ejection from the helical surface for $\lambda/2R=0.9$, and volume $>$ 20 $\upmu$l. For an Eruciform drop, the oscillations are small, but for a spherical shape, the oscillations are larger than the diameter of the helix. \par


\noindent\textbf{SI10: Renolds number and Capillary number:}
The Reynolds number ($Re=\rho v g/\upmu$, where $\rho$, $v$, $\upmu$ are density, downward velocity, and dynamic viscosity of the drop, respectively) calculated for flow regimes shown in Figure S9(a). For some cases (larger pitch and larger drop volume) the $Re>1$, so the inertia will affect the flow of a drop on a fiber, which is a possible cause of deviation from the scaling relation for velocity. Low values of capillary number ($Ca=v \upmu/\gamma$, where $\gamma$ is the surface tension) represent that the drop flow on the helical fiber is affected by the surface tension of the liquid used (Figure S9(b)).
 \begin{figure}[htbp]
	\centering
		\includegraphics[width=0.8\textwidth]{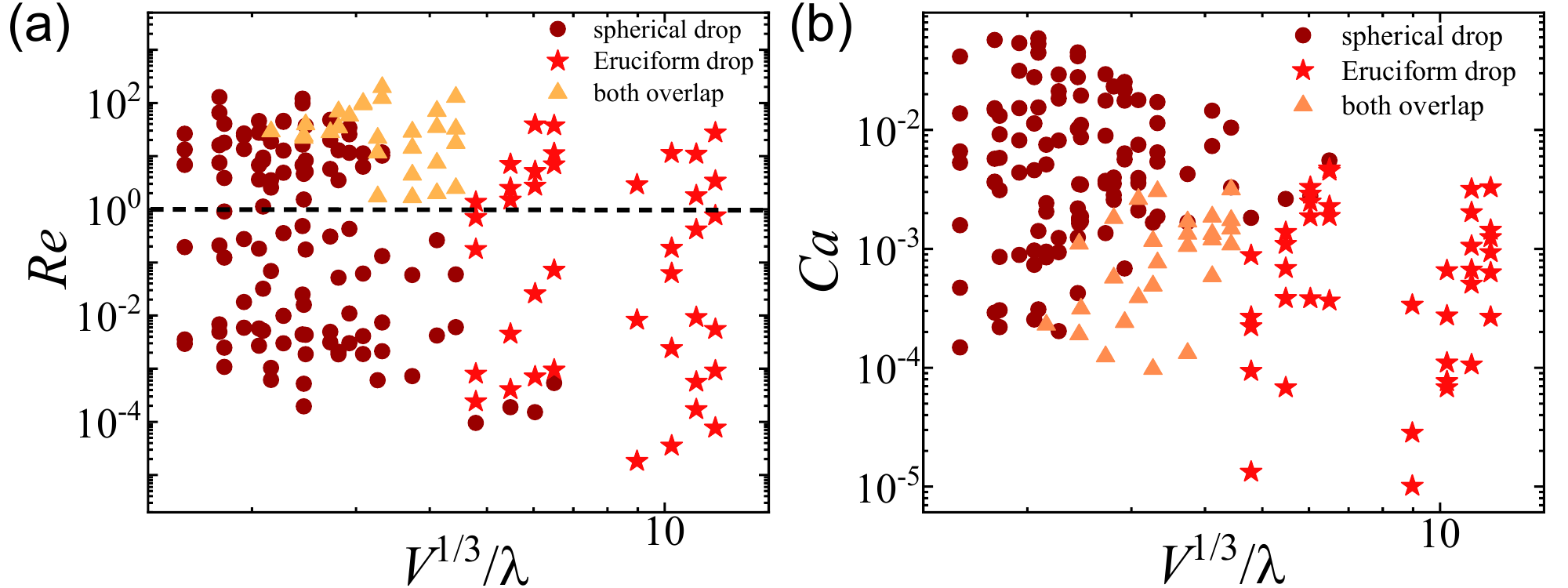}
	\caption{(a),(b) Reynolds number and capillary number for all viscosity ranges from 0.001 Pa$\cdot$s to 1.412 Pa$\cdot$s.}
	\label{FigureS10}
\end{figure}

\noindent\textbf{SI11: Wall adhesion for larger pitch values:}
\begin{figure}[htbp]
	\centering
		\includegraphics[width=0.85\textwidth]{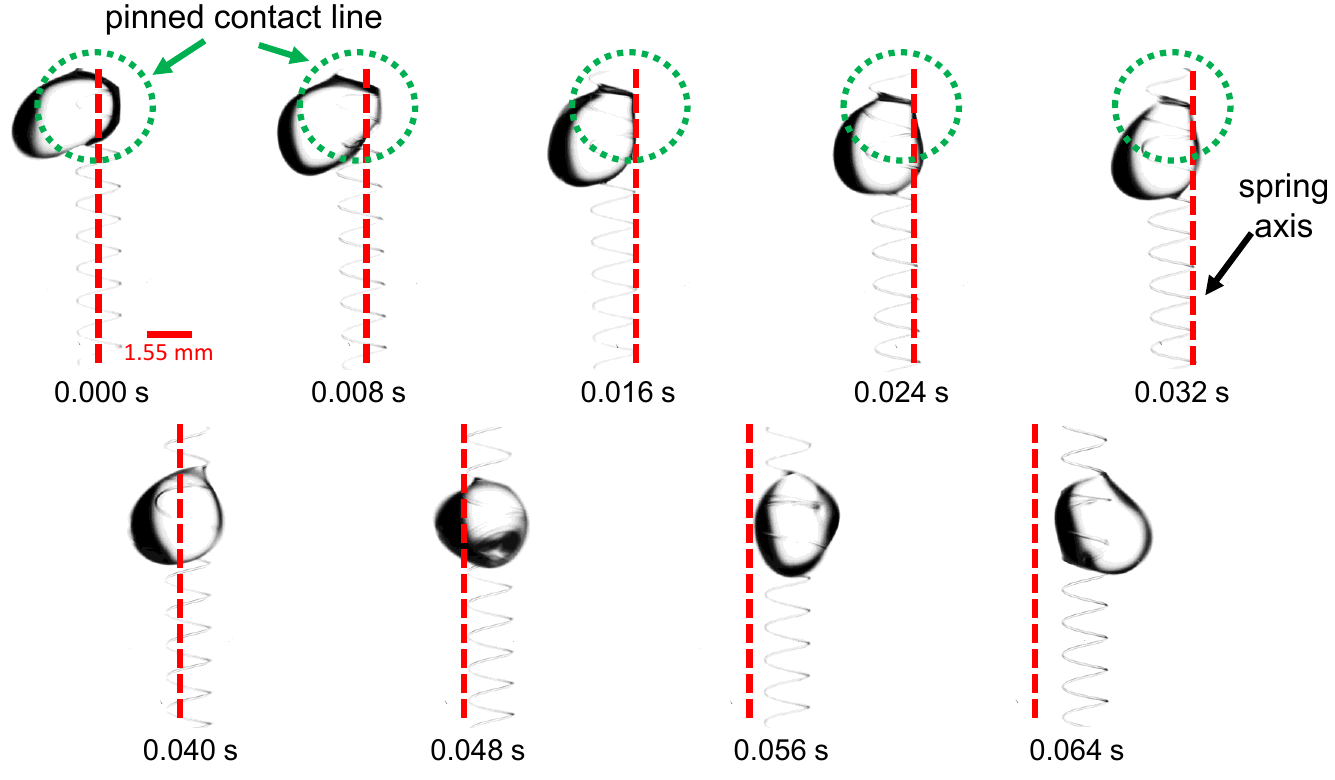}
	\caption{Drop experiencing an adhesion force with the helical spring because its size is larger than the diameter of the spring (inside red dotted circle). The scale bar is the same for all images.}
	\label{FigureS11}
\end{figure}

\noindent\textbf{SI12: Drop instability while flowing in a helical spring:}
\begin{figure}[htbp!]
	\centering
		\includegraphics[width=0.8\textwidth]{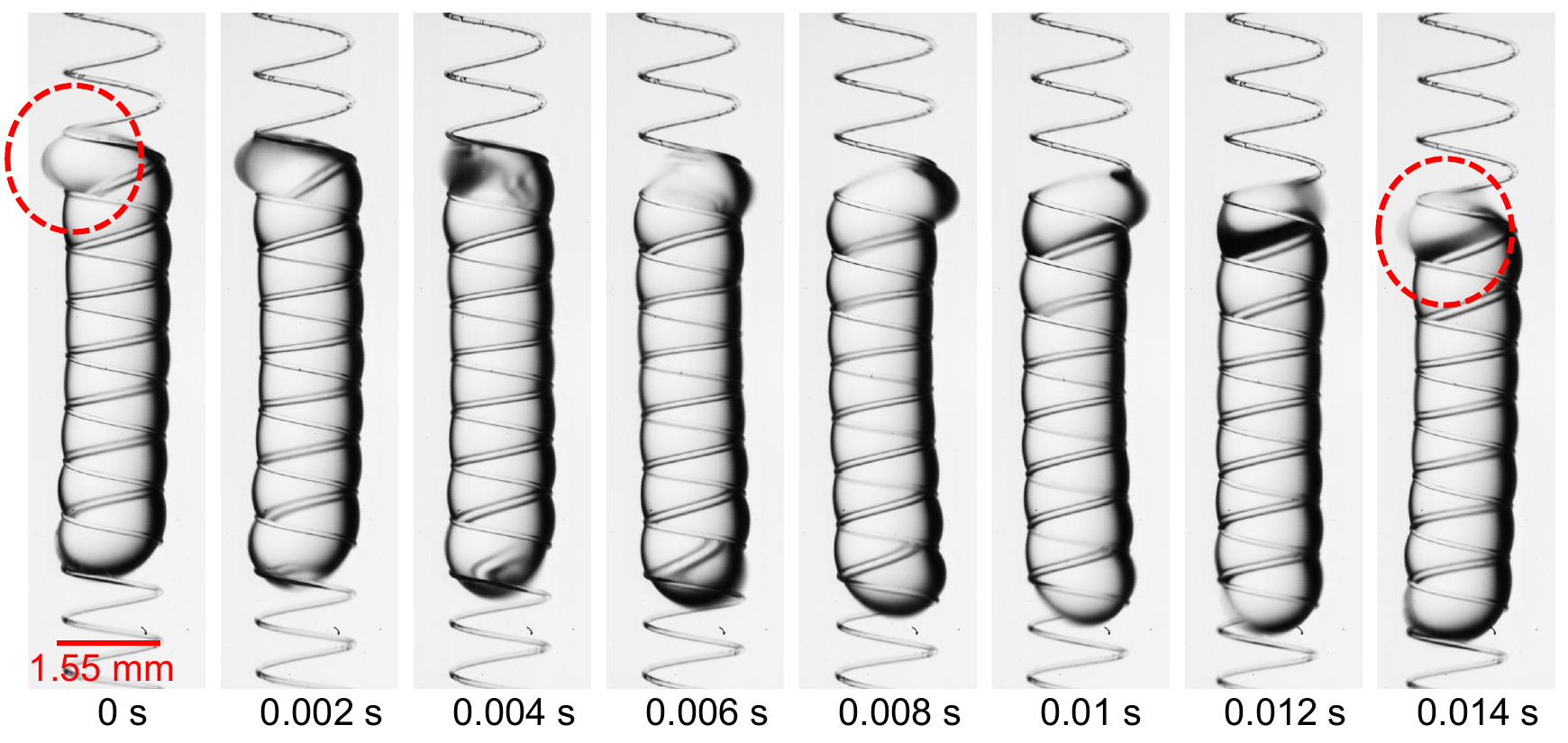}
	\caption{Small droplets ejection from the back receding end of the drop while moving on a helical spring. The scale bar is the same for all images.}
	\label{FigureS12}
\end{figure}
For an Eruciform droplet $>$ 30 $\upmu$l volume, the drop begins to bulge out from different parts of the droplet. If the pitch of the helix is small enough to hold the droplet in an Eruciform shape, then the drop shows the small droplets ejecting out from the back side of the droplet, indicated inside a red dotted circle in Figure S11. \par


\noindent\textbf{SI13: Viscosity of the glycerol/water solution:}
\begin{table}[htbp]
\centering
\begin{tabular}{|c|c|} 
 \hline
  Liquid (w/w $\%$) &  Viscosity (Pa$\cdot$s) at 20 $^\circ$C \\ [0.5ex] 
 \hline\hline
Water : Glycerol (100:00)
 & 0.001\\
 \hline
Water : Glycerol (70:30)
 & 0.0025\\
 \hline
 Water : Glycerol (50:50)
 & 0.006\\
 \hline
Water : Glycerol (40:60)
 & 0.0108\\ 
 \hline
 Water : Glycerol (15:85)
 & 0.109\\ 
\hline
 Water : Glycerol (04:96)
 & 0.624\\ 
\hline
 Water : Glycerol (0:100)
 & 1.412\\ 
 \hline
\end{tabular}
\caption{Viscosity of water-glycerol solution at 20 $^\circ$C \cite{segur1951viscosity}}
\label{TableS1}
\end{table}
A water-glycerol solution is used to achieve a viscosity range from 0.001 Pa$\cdot$s (100$\%$ water) to 1.412 Pa$\cdot$s (100$\%$ glycerol). The water and glycerol were mixed in different weight $\%$  and their corresponding viscosity value are shown in Table S1.

\end{suppinfo}

\end{document}